# Lipid Bilayer-Coated Curcumin-based Mesoporous Organosilica Nanoparticles for Cellular Delivery

Stefan Datz,[a] Hanna Engelke,[a] Constantin v. Schirnding,[a] Linh Nguyen,[a] Thomas Bein*[a]

**Effective and controlled drug delivery systems with on-demand release abilities and biocompatible properties receive enormous attention for biomedical applications. Here, we describe a novel inorganic-organic hybrid material with a strikingly high organic content of almost 50 wt%. The colloidal periodic mesoporous organosilica (PMO) nanoparticles synthesized in this work consist entirely of curcumin and ethane derivatives serving as constituents that are crosslinked by siloxane bridges, without any added silica. These mesoporous curcumin nanoparticles (MCNs) exhibit very high surface areas (over 1000 m$^2$/g), narrow particle size distribution (around 200 nm) and a strikingly high stability in simulated biological media. Additionally, the MCNs are used as a cargo delivery system in live-cell experiments. A supported lipid bilayer (SLB) efficiently seals the pores and releases Rhodamin B as model cargo in HeLa cells. This novel nanocarrier concept provides a promising platform for the development of controllable and highly biocompatible theranostic systems.**

## Introduction

Periodic mesoporous organosilica (PMO) constitutes a new type of inorganic-organic porous hybrid material, which holds great promise in a variety of fields such as chemical sensing,[1-7] catalysis[8-12] and biomedical applications.[13-15] Since the independent discovery of this new class of mesoporous materials in the groups of Inagaki, Stein and Ozin in 1999,[16-18] PMO materials, synthesized by using bridged silsesquioxanes as precursors, have recently been prepared at the nanoscale.[19-21] Different approaches were used to synthesize PMO nanoparticles with simple, low-molecular-weight organosilane bridging groups. In a sol-gel process using Pluronic P123 as the template, Landskron et al. synthesized rodlike nanoparticles with adjustable aspect ratios.[22] Using cetyltrimethylammonium bromide (CTAB) as the micellar template and an ammonia-catalyzed sol-gel reaction, Huo et al. prepared highly ordered and dispersable PMO nanoparticles with methane, ethane, ethylene and benzene organic bridging groups within the pore walls.[23] In another approach the group of Shi et al. used silica-etching chemistry to obtain hollow PMO nanoparticles that were used for nano-biomedical applications for the first time.[24] Recently, the group of Durand reported the synthesis of biodegradable PMO nanospheres and nanorods with a disulfide-containing organic bridging group. The morphology and size of these nanostructures was controlled by adjusting the ratio of bis(triethoxysilyl)ethane and bis(3-triethoxysilyl-propyl)-disulfide.[25] These mixed PMO nanospheres and rods were used as a biodegradable nanocarrier for doxorubicin in breast cancer cell lines. In the group of Kashab et al., enzymatically degradable silsesquioxane nanoparticles were synthesized and used as fluorescent nanoprobes for in vitro imaging of cancer cells.[26] Zink and co-workers developed different light-activatable and pH-responsive hybrid materials for drug delivery applications.[27-29] In these studies mostly low-molecular weight organic silsesquioxane bridging groups were incorporated into the pore walls of mesoporous nanostructures. Here, we report the synthesis of a PMO nanomaterial consisting of the biocompatible and large molecule curcumin and ethane organic moieties without the use of additional silica.

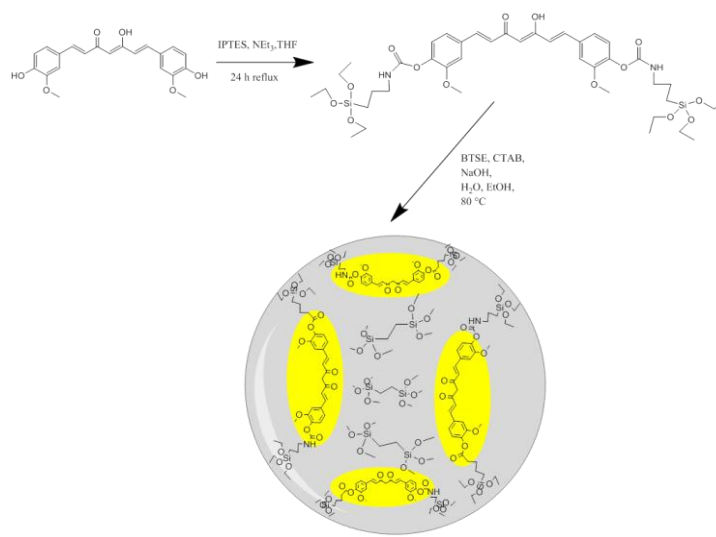

**Fig. 1** Schematic representation of the synthesis and the inorganic-organic hybrid composition of yellow-colored mesoporous curcumin nanoparticles (MCNs)

Curcumin is a natural yellow-colored antioxidant compound extracted from *Curcuma longa* and has been used for centuries in its crude form as dietary supplement and in traditional Asian medicines.[30] Recently, it has been shown that curcumin exhibits an exceptionally large range of biomedical activity against diseases such as Alzheimer, Parkinson, Malaria and many more.[31] In addition, it shows strong anti-inflammatory effects and has potential chemotherapeutic value as it inhibits cell proliferation and induces apoptosis in various cancer cell lines.[32-35] However, its bioavailability is limited by its very low aqueous solubility.[36, 37] Many different approaches have been investigated to improve the bioavailability and biopharmaceutical properties such as incorporating curcumin into liposomes,[38, 39] polymeric nanoparticles[40-42] or amino acid conjugates.[43, 44] Various successful *in vitro*[45-47] and *in vivo*[48-50] studies show the exceptional anticancer properties of curcumin nanoformulations. Additionally, it is well tolerated by the human body up to 12 g/day in oral administration as shown in clinical studies, which shows great promise regarding the biocompatibility of curcumin-based nanosystems.[51] Here, we present the synthesis of PMO nanoparticles with curcumin being the main organic constituent of the organosilica framework. Importantly, the synthesis was achieved without the addition of tetraethyl orthosilicate (TEOS), which is often used in other PMO studies for framework stabilization. The nanoparticles obtained in this study exhibit good dispersibility and high porosity parameters, which hold promise for a variety of applications in drug delivery. Furthermore, the incorporated curcumin compounds cause significant fluorescence of the nanoparticles themselves, which implies that no additional dye is necessary to track the NPs in live-cell experiments. The mesoporous PMO nanoparticles were used as a cargo release system with a Supported Lipid Bilayer (SLB) serving as cap in various *in vitro* experiments.

## Results and Discussion

The new nanomaterial was synthesized starting with the preparation of precursors. The synthesis of the precursor curcumin-IPTES was achieved following a previously described procedure.[5]

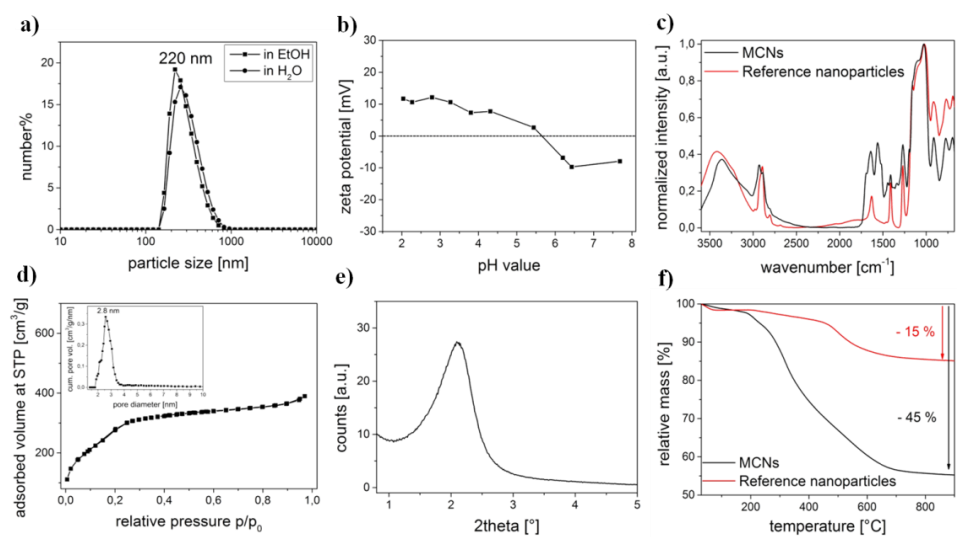

**Fig. 2** Characterization of the Curcumin PMO nanoparticle material. a) DLS measurements in ethanol and water, b) zeta potential measurement, c) infrared spectrum, d) nitrogen sorption isotherm and pore size distribution (inset), e) small-angle X-ray scattering (SAXS), f) thermogravimetric analysis of MCNs and reference nanoparticles without curcumin.

In this reaction, curcumin and 3-isocyanatopropyl-(triethoxy)silane (IPTES) form carbamate linkages under anhydrous basic conditions. The completion of the reaction can be monitored by infrared and UV-VIS spectroscopy in addition to NMR data shown in the experimental section. The UV-VIS spectrum (see SI, Figure S1) shows a significant blue-shift in the absorption from 430 to 415 nm after addition of the electron-withdrawing carbamate-linked silyl groups next to the conjugated π-electron system of the curcumin compound. In the IR spectra, the completion of the reaction can be followed by the disappearance of the characteristic isocyanate vibration at 2270 $cm^{-1}$ and the increasing intensity of the C=O stretching vibration due to the carbamate linkage group absorbing at 1710 $cm^{-1}$. The synthesized precursor was then used in a carefully controlled sol-gel reaction to form mesoporous curcumin nanoparticles (MCNs). The synthesis was performed in a water-ethanol mixture with cetyltrimethylammonium bromide as the micellular template under slightly basic conditions (Figure 1). The addition of ethanol was crucial because of the low solubility of Curcumin-IPTES in aqueous solutions. The template preparation, the catalyzed hydrolysis of Curcumin-IPTES and BTSE acting as the silica sources, and the nanoparticle formation was performed at 80 °C. After completion of the reaction the template was extracted in an ammonium nitrate containing ethanolic solution followed by an additional extraction step with ethanol under reflux. After several washing steps and redispersion in absolute ethanol the synthesis resulted in a colloidal yellow suspension of MCNs (see SI, Figure S2b). Dynamic light scattering (DLS) measurements showed a narrow size distribution of MCNs with a maximum around 220 nm (Figure 2a), implying excellent colloidal stability in ethanol and in water. Compared to silica particles, the zeta potential measurement of MCNs shows an increased isoelectric point at pH 5.5, which is due to the strongly reduced amount of negatively charged silanol groups on the surface of the nanoparticles, compared to common mesoporous silica nanoparticles (Figure 2b).[52] In order to investigate the total organic amount within the PMO framework, a reference PMO nanoparticle material was synthesized. These reference nanoparticles consist exclusively of ethane groups as the organic linker in the PMO material.

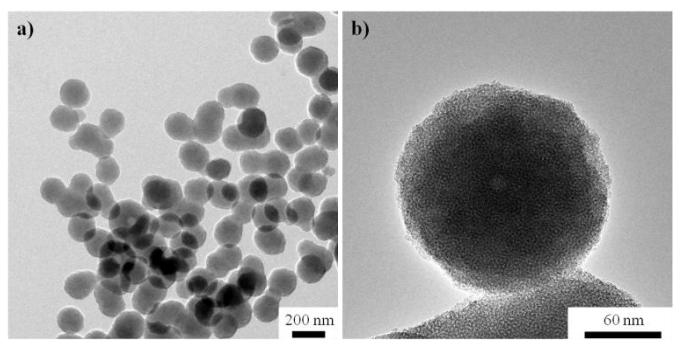

**Fig. 3**  Transmission electron microscopy characterization of the MCNs.

IR data for both types of samples depicted typical vibrational modes of the silica framework between 780 and 1300 $cm^{-1}$ (Figure 2c). The shoulders at 1705 $cm^{-1}$ and 1510 $cm^{-1}$ in the MCN sample can be assigned to the stretching vibrations of C=O and N-H of the carbamate group, respectively. The peak at 1560 $cm^{-1}$ is attributed to the secondary amine vibrational modes. The intensive signal at 1640 $cm^{-1}$ is due to physisorbed water and can be seen in all spectra. The signals beyond 2800 $cm^{-1}$ are assigned to C-H and N-H stretching vibrations of the incorporated organic moieties. Nitrogen sorption and small-angle X-ray scattering (SAXS) measurements were used to characterize the porosity parameters of the obtained nanoparticles. The nitrogen sorption data (Figure 2d) showed a type IV isotherm with a strikingly high calculated BET surface area of 1040 $m^2/g$ and a narrow pore size distribution around 2.8 nm. The resulting pore volume is 0.55 $cm^3/g$. In SAXS measurements, only the first (100) reflex is observed, indicating a disordered worm-like pore structure of the mesopores in the nanoparticles (Figure 2e). Thermogravimetric analysis (TGA) data (Figure 2f) indicate a relative mass loss of 15 wt% up to 900 °C in the reference material consisting of a PMO with ethane groups but without curcumin. In comparison, the mixed PMO MCN nanoparticles containing curcumin and ethane organic bridging groups reveal a significantly enhanced mass loss of 45 wt% indicating the successful incorporation of curcumin into the silica framework. Raman spectroscopy and additional solid-state NMR spectra show specific signals for curcumin and ethane in the nanoparticle material as well (see SI, Figure S2a and S3). The $^{29}Si$-spectrum shows the presence of T-type signals between -45 and -70 ppm. The signals arising from the organic parts indicate that the organic linkers are completely included into the hybrid silica framework with a high degree of silanol condensation. Importantly, the absence of Q-type $^{29}Si$-signals provides evidence that the Si-C bonds are stable under the applied synthetic conditions. Electron microscopy was used to investigate the morphology, pore structure and size distribution of MCNs. TEM images of MCNs are depicted in Figure 3 and display spherically shaped particles with a very narrow particle size distribution. A radially disposed worm-like structure of the mesopores of MCNs can be seen in the STEM image (Figure S4c). 2D Fourier transformation of the image (FFT, Figure S4g) reveals a pore-to-pore distance of about 4.5 nm in good accordance to previously described sorption and SAXS measurements. MCNs show significant autofluorescence with an excitation maximum at about 400 nm and an emission maximum at about 520 nm (Fig. S7). This key feature allows us to observe them in a fluorescence microscope without addition or attachment of further dyes. The biological stability of MCNs was investigated in simulated body fluid (SBF) to gain insights regarding the reactivity of these particles in simulated biological media for future drug delivery applications. The long-term stability was investigated with nitrogen sorption, X-ray analysis, electron microscopy, and infrared spectroscopy (see SI, Figure S5 and S6). Strikingly, the particles were stable throughout the complete experimental time (up to 28 days) in SBF solution and showed no phase transition or crystallization behavior at all. In comparison, common mesoporous silica nanoparticles (MSNs) start to show degradation and formation of apatite-like structures after a few hours.[53] Our findings can be explained by the drastically decreased amount of reactive silanol groups and siloxane bridges on the surface of organosilica nanoparticles compared to classical silica-based MSNs. This feature is anticipated to make colloidal MCNs promising candidates for drug delivery applications where enhanced stability is desired. The highly porous and colloidal MCN nanoparticles were also investigated in live-cell experiments. Cellular uptake of MCNs and the release of model cargos were investigated in HeLa cells. When coated with a lipid bilayer, colloidal curcumin nanoparticles are internalized with high efficiency by HeLa cells after 24 h of incubation and can be detected based on their autofluorescence (Fig. S7). The release behavior of cargo-loaded MCNs was investigated *in vitro*. Rhodamin B - a dye that stains mitochondria in cells - was chosen as a model cargo and the particles were sealed with a supported lipid bilayer (SLB). The SLB was produced in a two-step approach employing first DOTAP only, followed by a DOTAP/DOPC mixture. The particles loaded with Rhodamin B and coated with the lipid layer were efficiently internalized by HeLa cells. After 24 h a slight release of Rhodamin B could be observed. Addition of chloroquine leads to an enhanced release after 48 h as shown in Figure 4.

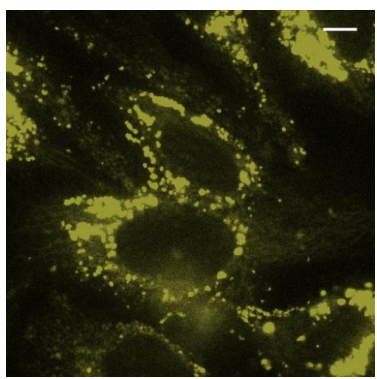

**Fig. 4** Rhodamin B-loaded MCNs in HeLa cells. Release can be observed after 48 h incubation and addition of chloroquine. Scale bar: 10 μm.

## Conclusions

To conclude, we have developed new colloidal periodic mesoporous organosilica nanoparticles containing curcumin as organic component, with very high organic wall content. They are fluorescent, possess a large pore volume and surface area and show very high stability in simulated body fluid. When coated with a lipid layer they are successfully internalized by cells and can deliver and release Rhodamin B in those cells. Thus, they show great potential for future applications as drug delivery system.

## ACKNOWLEDGMENT

The authors acknowledge financial support from the Deutsche Forschungsgemeinschaft (DFG) in the context of SFB 749 and SFB 1032, the Excellence Clusters Nanosystems Initiative Munich (NIM) and Center for Integrated Protein Science Munich (CIPSM), and the Center for Nano Science (CeNS). We thank Prof. Christoph Bräuchle (LMU) for fruitful discussions and for giving us access to his spinning disc microscope. We thank Dr. Markus Döblinger, Dr. Steffen Schmidt and Dr. Andreas Wisnet for electron microscopy.

**Supporting Information**

**Experimental section**

**Materials.** Curcumin (60-70%), 3-isocyanatopropyl(triethoxysilane) (IPTES), tetrahydrofuran (dry), triethylamine (97%), cetyltrimethylammonium bromide, ammonium nitrate, ammonium bicarbonate, Rhodamin B, calcein, sodium hydroxide, DMSO-$d_6$, dichloromethane, CDCl$_3$, methanol were purchased from Sigma Aldrich. Bis(triethoxysilyl)ethane (BTSE) was purchased from ABCR. DOPC (1,2-dioleoyl-*sn*-glycero-3-phosphocholine) and DOTAP (1,2-dioleoyl-3-trimethylammonium propane) were purchased from Avanti Polar Lipids. All chemicals were used as received without further purification. Doubly distilled water from a Millipore system (Milli-Q Academic A10) was used for all synthesis and purification steps.

**Synthesis of Curcumin-Precursor.** Curcumin ((1*E*,6*E*)-1,7-bis-(4-hydroxy-3-methoxy-phenyl)-hepta-1,6-dien-3,5-dion, 2.00 g, 5.43 mmol, 1 eq.) was dissolved in 25 mL dry THF in a three-necked flask. Subsequently, 3-isocyanatopropyl(triethoxysilane) (5.37 g, 21.7 mmol, 4 eq.) and triethylamine (165 µg, 1.63 mmol, 0.3 eq.) were added under stirring and the mixture was refluxed for 24 h at 85 °C in a nitrogen flow. After cooling down to room temperature, the sample was filtered and washed with ethyl acetate. The solvents were evaporated at reduced pressure and the sample was purified with column chromatography on silica gel with a solvent mixture containing 97 v% dichloromethane, 2 v% methanol and 1 v% triethylamine. The compound obtained (named Curcumin-IPTES) was dried under high vacuum for 12 hours and used without further purification (yield: 2.29 g, 2.65 mmol, 49 %). $^1$H-NMR (300 MHz, DMSO-$d_6$): δ [ppm] = 7.72 (t, 2H), 7.62 (d, 2H), 7.44 (s, 2H), 7.26 (d, 2H), 7.09 (d, 2H), 6.95 (d, 2H), 6.16 (s, 1H), 3.80 (s, 6H), 3.73 (qa, 18H), 1.45 (q, 6H), 1.12 (t, 27H), 0.55 (t, 6H). $^{13}$C-NMR (400 MHz, CDCl$_3$): δ [ppm] = 183.14, 153.98, 152.00, 141.75, 140.10, 133.36, 124.00, 123.67, 121.10, 111.47, 101.68, 58.49, 51,92, 43.72, 23.61, 18.29, 14.20. MS (ESI): [M-H]$^-$ cal.: 861.36206, found: 861.36672.

**Synthesis of mesoporous Curcumin-PMO nanoparticles (MCNs).** In a two-step sol-gel reaction, cetyl trimethyl-ammonium bromide (CTAB, 0.96 mmol, 350 mg) was dissolved in a mixture containing 120 mL H$_2$O and 15 mL absolute ethanol in a 250 ml round bottom flask. Subsequently, 875 µL sodium hydroxide solution (2 M) was added and the mixture was stirred at 80 °C for 30 minutes. In a glass vessel, 400 mg Curcumin-IPTES (0.32 mmol) was mixed with 200 µL bis(triethoxysilyl)ethane (BTSE, 0.51 mmol) and 400 µL ethanol. This precursor solution was preheated to completely dissolve the compounds and afterwards quickly injected into the stirring aqueous template solution. The reaction was maintained for 90 minutes at 80 °C and 700 rpm. Extraction of the organic template was achieved by heating the ethanol-suspended (80 mg) sample under reflux at 90 °C for 1 h in a mixture of 2 g ammonium nitrate and 100 mL ethanol. Afterwards, the sample was centrifuged for 15 minutes at 7830 rpm (7197 rcf), redispersed in ethanol and heated under reflux at 90 °C in a solution of 100 mL ethanol for 45 minutes. After centrifugation, the particles were re-dispersed in 20 mL ethanol, resulting in a colloidal yellow suspension with a concentration of 4 mg/mL.

**Degradation study in Simulated Body Fluid (SBF).** To prepare the SBF buffer, the following reagents were dissolved in bi-distilled water and the solution was filled up to 1000 ml: 6.057 g NH$_2$C(CH$_2$OH)$_3$ (TRIS), 0.350 g NaHCO$_3$, 0.224 g KCl, 7.996 g NaCl, 0.228 g K$_2$HPO$_4$·3 H$_2$O, 0.305 g MgCl$_2$·6 H$_2$O, 0.278 CaCl$_2$, 0.071 g Na$_2$SO$_4$. The pH of the solution was adjusted to 7.4 at 37 °C with 1 M HCl. 100 mg of as-synthesized Curcumin nanoparticles were centrifuged, washed with bi-distilled water and redispersed in 50 mL SBF buffer solution. The mixture was stored at 37 °C and at selected times, 10 mL of the solution containing nanoparticles was collected, centrifuged and washed with water for several times to remove deposited salts. The sample was then dried at 70 °C and used for further characterization. The collected supernatants were used for pH measurements. The selected times to collect the samples were 3 h, 1, 4, 7 and 28 days.

**Lipid preparation.** The following lipids were used: DOPC (1,2-dioleoyl-*sn*-glycero-3-phosphocholine, Avanti Polar Lipids), DOTAP (1,2-dioleoyl-3-trimethylammonium propane, Avanti Polar Lipids). The amount of 2.5 mg of the individual lipids was dissolved in a 1 mL mixture of 40 %vol absolute ethanol and 60 %vol MQ water (conc. 2.5 mg/mL).

**Cargo loading and Supported Lipid Bilayer-capping (SLB).** The amount of 0.5 mg of MCNs in ethanolic solution were centrifuged (4 min, 8609 rcf, at 15 °C) and redispersed in a 1000 µL loading mixture containing Rhodamin b (0.5 mM in water). The particles were centrifuged after 2 h of loading (4 min, 8609 rcf, at 15 °C), separated from the loading solution and 100 µL of the above DOTAP solution was added. Upon addition of 900 µL MQ water (pH adjusted to 9.4 with sodium hydroxide) the formation of the first SLB layer on the external surface of MCNs was induced. After centrifugation (4 min, 8609 rcf, at 15 °C) and redispersion in 100 µL of a 1:1 mixture of the above DOPC/DOTAP solutions, the formation of a second layer around the MCNs was induced by adding 900 µL HBSS buffer.

**Characterization.** All samples were investigated with an FEI Titan 80-300 transmission electron microscope operating at 300 kV with a high-angle annular dark field detector. A droplet of the diluted MSN solution in absolute ethanol was dried on a carbon-coated copper grid. Dynamic light scattering (DLS) measurements were performed on a Malvern Zetasizer-Nano instrument equipped with a 4 mW He-Ne laser (633 nm) and an avalanche photodiode. The hydrodynamic radius of the particles was determined by dynamic light scattering in ethanolic suspension. For this purpose, 100 µL of an ethanolic suspension of MSN (ca. 10 mg/mL) was diluted with 3 mL of ethanol prior to the measurement. Zeta potential measurements of the samples were performed on a Malvern Zetasizer-Nano instrument equipped with a 4 mW He-Ne laser (633 nm) and an avalanche photodiode. Zeta potential measurements were performed using the add-on Zetasizer titration system (MPT-2) based on diluted NaOH and HCl as titrants. For this purpose, 1 mg of the particles was diluted in 10 mL bi-distilled water. Nitrogen sorption measurements were performed on a Quantachrome Instruments NOVA 4000e. All samples (10 mg each) were heated to 100 °C for 12 h in vacuum (10 mTorr) to outgas the samples, before nitrogen sorption was measured at 77 K. Pore size and pore volume were calculated with a NLDFT adsorption branch model of $N_2$ on silica, based on the adsorption branch of the isotherms. A BET model was applied in the range of 0.05 – 0.20 $p/p_0$ to evaluate the specific surface area of the samples. Centrifugation was performed using an Eppendorf centrifuge with an adapter for Falcon tubes or an Eppendorf centrifuge 5418 for small volumes. Raman spectra were recorded on a Jobin Yvon Horiba HR800 UV Raman microscope using a He-Ne laser emitting at λ = 633 nm with a laser power of 10 mW. IR measurements were performed on a Bruker Equinox 55 FTIR spectrometer in absorbance mode (spectra were background substracted). UV-VIS spectra were recorded with a NanoDrop ND 1000 spectrometer. Usually, 2 µL of sample were used and all presented spectra are background corrected for water absorption. Thermogravimetric analysis (TGA) of the samples (about 10 mg of dried nanoparticles) was performed on a Netzsch STA 440 Jupiter thermobalance with a heating rate of 10 K/min in a stream of synthetic air of about 25 mL/min. Cross-polarized $^{29}$Si- and $^{13}$C-MAS NMR measurements were performed on a Bruker DSX Avance500 FT spectrometer in a 4 mm $ZrO_2$ rotor. The spinning rate was 6 kHz and a total number of 256 scans were recorded.

**Cell culture.** HeLa cells were grown in DMEM supplemented with 10 % FBS (lifetechnologies) at 37 °C in a humidified atmosphere containing 5 % $CO_2$. They were seeded into ibiTreat 8 well slides (ibidi GmbH) at concentrations of 5000 – 10 000 cells per well the day prior to treatment. For internalization as well as release studies 2 µg of particles were added per well. For internalization studies the cell membranes were stained immediately before imaging by addition of wheat germ agglutinin Alexa Fluor 647 conjugate (WGA 647, lifetechnologies) at a final concentration of 5 µg/mL and subsequent washing with DMEM medium.

**Live cell imaging.** Cells were imaged 24 or 48 h after incubation with particles on a spinning disc microscope (Zeiss Cell Observer SD utilizing a Yokogawa spinning disk unit CSU-X1). The objective was a 63x plan apochromat oil immersion objective (NA 1.4, Zeiss). The exposure time was 0.1 s. For MCN imaging the excitation was 488 nm, for Rhodamin B 561 nm and for WGA 647 633 nm laser light.

**Further Characterization**

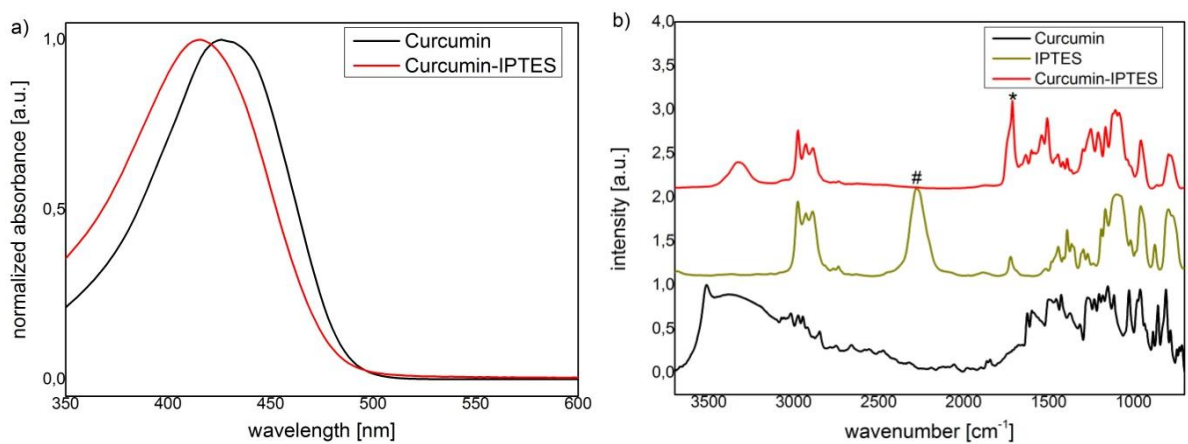

Figure S1: UV-VIS absorption (a) and IR spectroscopy (b) data on the formation of the precursor.

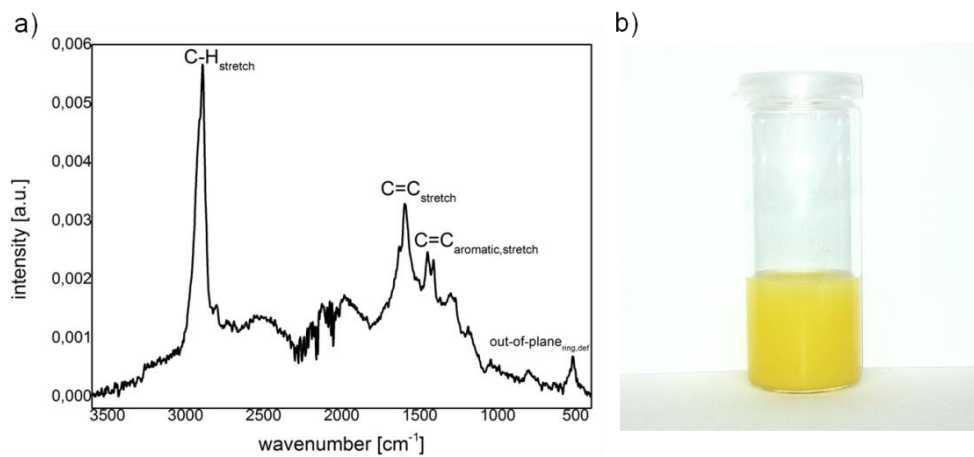

Figure S2: Raman spectrum of MCNs (a), image of colloidal MCNs stored in ethanol (b).

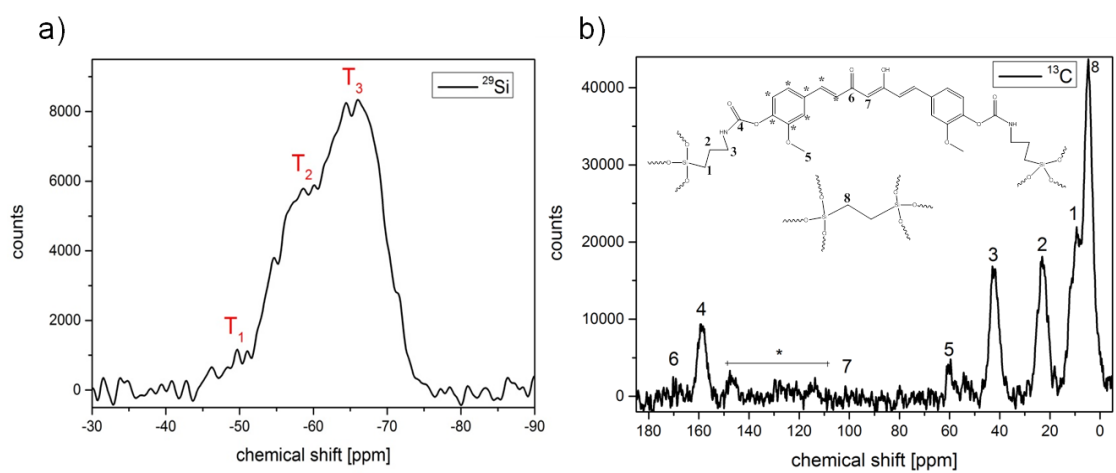

Figure S3: $^{29}$Si MAS ssNMR (a) and $^{13}$C MAS ssNMR (b) of MCNs.

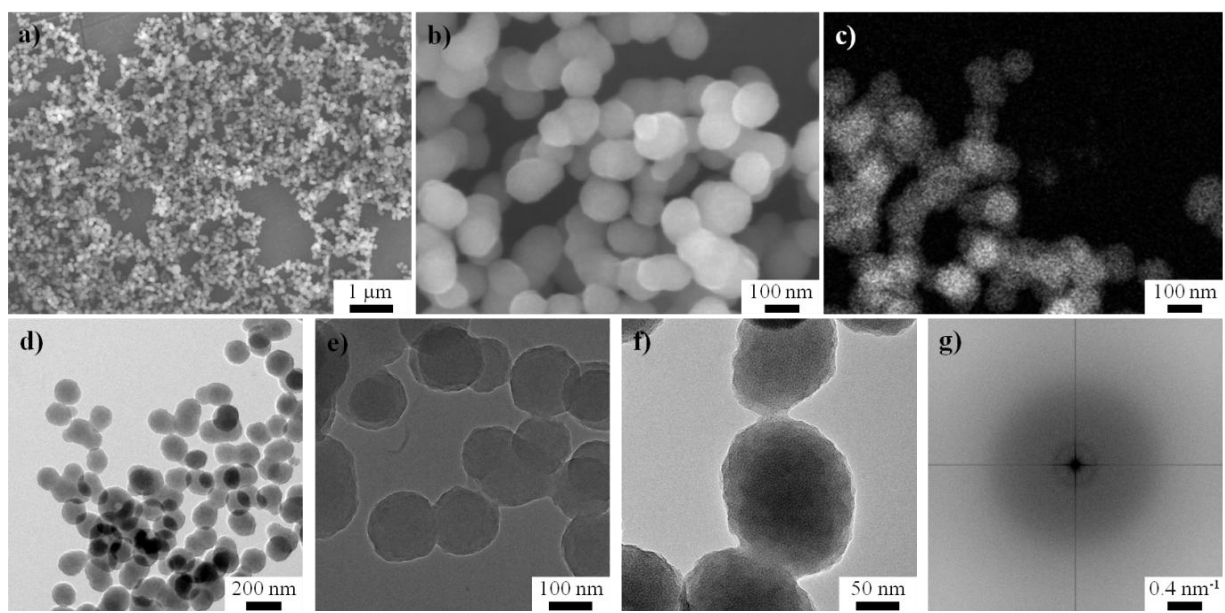

Figure S4: Electron microscopy of MCNs. SEM images (a, b), STEM image (c), TEM images (d-f), FFT (g).

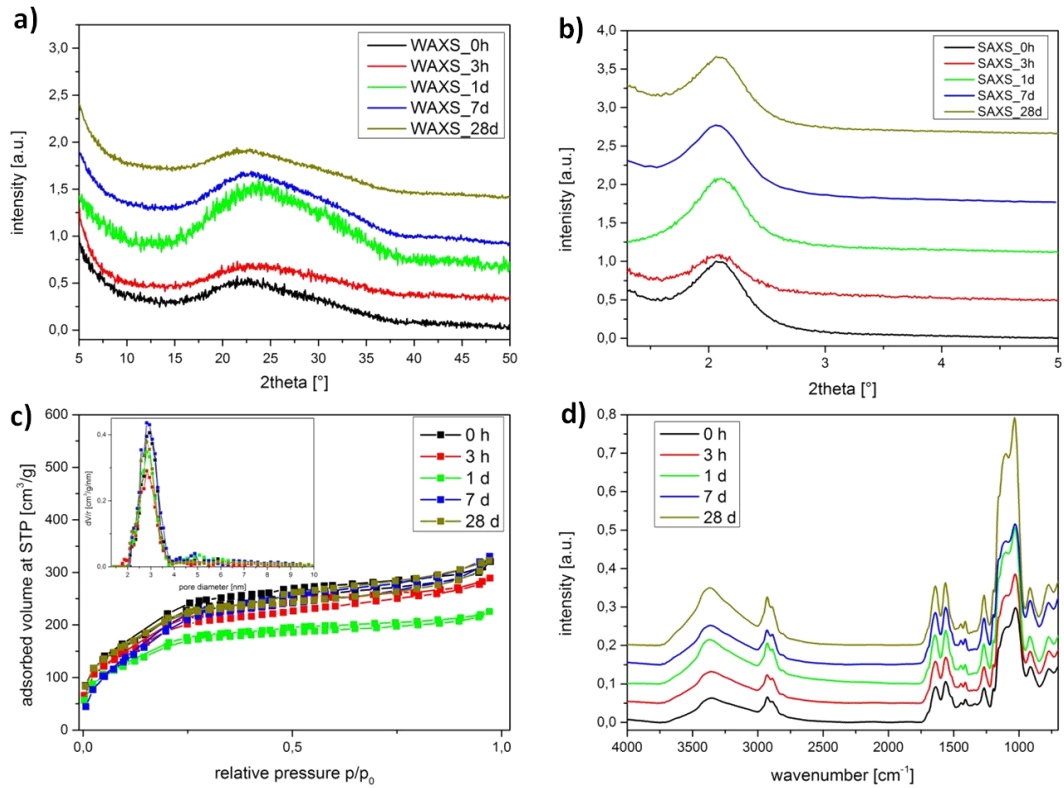

Figure S5: Biostability study. Wide-angle X-ray scattering data (a, WAXS), small-angle X-ray scattering data (b, SAXS), nitrogen sorption isotherm (c, inset: pore size distribution), infrared spectroscopy (d).

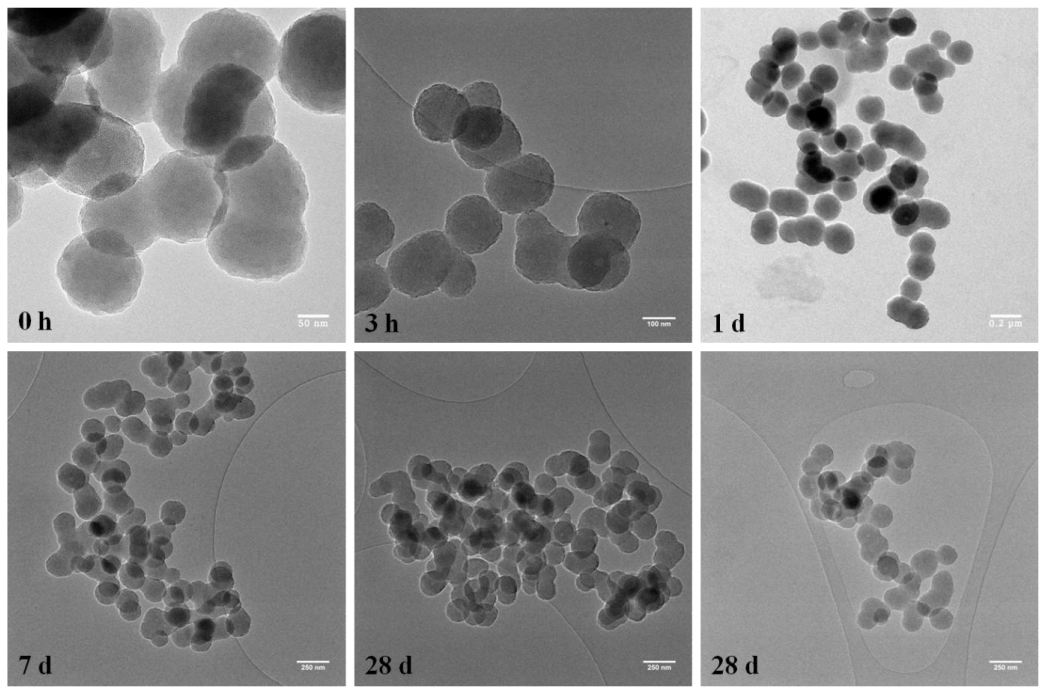

Figure S6: Transmission electron microscopy (TEM) images at different time points of MCNs in SBF (0 h, 3 h, 1 d, 7 d, 28 d) as part of the biodegradability test.

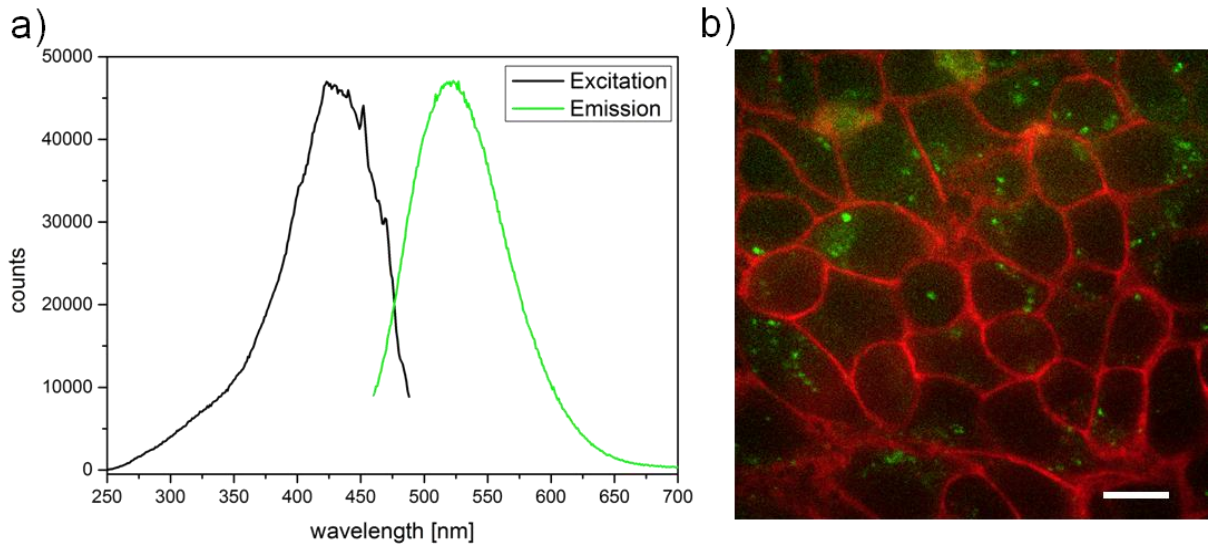

Figure S7: Fluorescence excitation and emission spectra of MCNs in PBS buffer (a). Cellular uptake (b) of DOTAP-coated MCNs (green) after 24 h incubation on HeLa cells (red: WGA647 membrane staining); scale bar represents 10 µm.

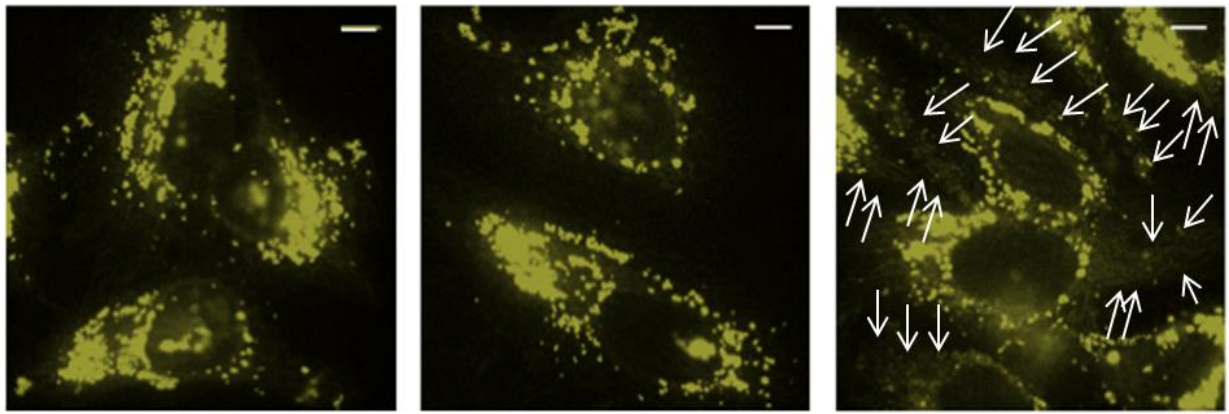

Figure S8: Spinning disc micrographs of Rhodamin B loaded MCNs internalized by HeLa cells. After 24 h (a) and 48 h (b) only a slight release of Rhodamin B is visible. Upon addition of chloroquine release is significantly **increased** (indicated by white arrows) (c). Scale bar: 10 µm.